\documentstyle[12pt]{article}
\title{Stochastic stability in spatial games} 
\author{Jacek Mi\c{e}kisz \\ Institute of Applied Mathematics \\
and Mechanics \\ Warsaw University  \\ ul. Banacha 2  \\ 02-097
Warsaw, Poland \\ e-mail: miekisz@mimuw.edu.pl} 
\begin{document} 
\baselineskip=20pt
\maketitle 

\noindent {\bf Abstract}: We discuss similarities and differences between systems 
of interacting players maximizing their individual payoffs and particles minimizing 
their interaction energy. Long-run behavior of stochastic dynamics of spatial games 
with multiple Nash equilibria is analyzed. In particular, we construct an example 
of a spatial game with three strategies, where stochastic stability of Nash equilibria 
depends on the number of players and the kind of dynamics.
\vspace{3mm}


\eject

\newtheorem{theo}{Theorem}
\newtheorem{defi}{Definition}
\newtheorem{hypo}{Hypothesis}

\section{Introduction}

\noindent Many socio-economic systems can be modeled as systems of interacting 
individuals; see for example Santa Fe collection of papers on economic complex systems 
\cite{santa} and econophysics bulletin \cite{ekono}. One may then try to derive 
their global behavior from individual interactions between their basic entities. 
Such approach is fundamental in statistical physics which deals with systems 
of interacting particles. We will explore similarities and differences between systems 
of interacting players maximizing their individual payoffs and particles minimizing 
their interaction energy. 

Here we will consider game-theoretic models of many interacting agents \cite{wei,hof2,ams}. 
In such models, agents have at their disposal certain strategies and their payoffs 
in a game depend on strategies chosen both by them and by their opponents. 
In spatial games, agents are located on vertices of certain graphs and they interact 
only with their neighbors \cite{blume1,ellis1,young2,ellis2,nowak1,nowak2,linnor,doebeli,
sabo,hauert}. The central concept in game theory is that of a Nash equilibrium. 
It is an assignment of strategies to players such that no player, for fixed strategies 
of his opponents, has an incentive to deviate from his curent strategy; 
the change can only diminish his payoff.

One of the best known game is that of a Prisoner's Dilemma game 
\cite{axelrod}. It has a unique Nash equilibrium, when both players defect. 
In fact, defection is the best response to both cooperation and defection of the opponent.
However, both players are better off when they cooperate. Dynamical aspects of spatial prisoner's 
dillema games were discussed in many papers, see for example 
\cite{nowak1,nowak2,linnor,doebeli,sabo,hauert}. Players in these games adapt 
to their environment by imitating those with biggest payoffs. It was shown that cooperation 
persists for a certain range of parameters. 

Here we will discuss games with multiple Nash equilibria. One of the fundamental problems 
in game theory is the equilibrium selection in such games.
One of the selection methods is to construct a dynamical system 
where in the long run only one equilibrium is played with a high frequency. 
John Maynard Smith \cite{maynard1,maynard2} has refined the concept of equilibrium to include 
the stability of Nash equilibria against mutants. He introduced the fundamental notion 
of an evolutionarily stable strategy. If everybody plays such a strategy, then the small number 
of mutants playing a different strategy is eliminated from the population. 
The dynamical interpretation of the evolutionarily stable strategy was later provided 
by several authors \cite{tayjon,hof,zee}. They proposed a system of differential replicator 
equations, which describe the time-evolution of frequencies of strategies and analyzed 
the asymptotic stability of Nash equilibria.
  
Here we will discuss a stochastic adaptation dynamics of a population 
with a fixed number of players. In discrete moments of times, 
players adapt to their neighbors by choosing with a high probability
the strategy which is the best response, i.e. the one which maximizes 
the sum of the payoffs obtained from individual games. With a small probability, representing 
the noise of the system, they make mistakes. We study the long-time behavior of such dynamics.
We say that a configuration of strategies is {\bf stochastically stable} \cite{foya} if it has 
a positive probability in the stationary state of the above dynamics in the zero-noise limit, 
that is zero probability of mistakes. It means that in the long run we observe it with 
a positive frequency.

In Section 2, we introduce basic notions of game theory and discuss similarities 
and differences between ground-state configurations in classical lattice-gas models 
and Nash configurations in game theory. Section 3 contains the description 
of a simple stochastic dynamics. In Section 4, we present an example 
of a spatial game with three strategies, where stochastic stability 
of Nash equilibria depends on the number of players and the kind of dynamics. 
Discussion follows in Section 5.

\section{Nash configurations}

To characterize a game-theoretic model one has to specify the set of players, 
strategies they have at their disposal and payoffs they receive.  
Although in many models the number of players is very large, 
their strategic interactions are usually decomposed into a sum of two-player games. 
Only recently, there have appeared some systematic studies of truly multi-player games 
\cite{kim,broom,multi}. Here we will discuss only two-player games with two or three strategies.
We begin with games with two strategies.
A generic payoff matrix is given by

\noindent {\bf Example 1}
\vspace{3mm}

\hspace{23mm} A  \hspace{2mm} B   

\hspace{15mm} A \hspace{3mm} a  \hspace{3mm} b 

U = \hspace{6mm} 

\hspace{15mm} B \hspace{3mm} c  \hspace{3mm} d,

where the $ij$ entry, $i,j = A, B$, is the payoff of the first (row) player when
he plays the strategy $i$ and the second (column) player plays the strategy $j$. 
We assume that both players are the same and hence payoffs of the column player are given 
by the matrix transposed to $U$; such games are called symmetric. 

An assignment of strategies to both players is a Nash equilibrium, if for each player, 
for a fixed strategy of his opponent, changing the current strategy will not increase his payoff.
If $c>a, d>b$ and $a>d$, then the game has a unique Nash equilibrium $(B,B)$ but both players 
are much better off when they play $A$ - 
this is the classic Prisoner's dilemma case \cite{axelrod}. 
If $a<c$ and $d<b$, then there are two nonsymmetric Nash equilibria: $(A,B)$ and $(B,A)$ 
(a Hawk-Dove game \cite{maynard2}). Below we will discuss games with multiple Nash equilibria. 
If $a>c$ and $d>b$, then both $(A,A)$ and $(B,B)$ are Nash equilibria. 
If $a+b<c+d$, then the strategy $B$ has a higher expected payoff against a player playing 
both strategies with the probability $1/2$. We say that $B$ risk dominates the strategy $A$ 
(the notion of the risk-dominance was introduced and thoroughly studied 
by Hars\'{a}nyi and Selten \cite{hs}). If at the same time $a>d$, 
then we have a selection problem of choosing between the payoff-dominant 
(Pareto-efficient) equilibrium $(A,A)$ and the risk-dominant $(B,B)$.

Let us now describe spatial games with local interactions.
Let $\Lambda$ be a finite subset of the simple lattice ${\bf Z}^{d}$.
Every site of $\Lambda$ is occupied by one player who
has at his disposal one of $k$ different strategies ($k=2$ in the above example). 
Let $S$ be the set of strategies, then $\Omega_{\Lambda}=S^{\Lambda}$ is the space
of all possible configurations of players, that is all possible assignments 
of strategies to individual players. For every $i \in \Lambda$, 
$X_{i}$ is the strategy of the $i-$th player in the configuration 
$X \in \Omega_{\Lambda}$ and $X_{-i}$ denotes strategies of all remaining players; 
$X$ therefore can be represented as the pair $(X_{i},X_{-i})$. 
Every player interacts only with his nearest neighbors and his payoff 
is the sum of the payoffs resulting from individual plays.
We assume that he has to use the same strategy for all neighbors. 
Let $N_{i}$ denote the neighborhood of the $i-$th player. 
For the nearest-neighbor interaction we have $N_{i}=\{j; |j-i|=1\}$,
where $|i-j|$ is the distance between $i$ and $j$.
For $X \in \Omega_{\Lambda}$ we denote by $\nu_{i}(X)$ the payoff 
of the $i-$th player in the configuration $X$:
\begin{equation}
\nu_{i}(X)=\sum_{j \in N_{i}}U(X_{i}, X_{j}),
\end{equation}
where $U$ is a $k \times k$ matrix of payoffs of a two-player symmetric game with $k$ strategies.

\begin{defi}
$X \in \Omega_{\Lambda}$ is a {\bf Nash configuration} if for every $i \in \Lambda$
and $Y_{i} \in S$, $\nu_{i}(X_{i},X_{-i}) \geq \nu_{i}(Y_{i},X_{-i})$.
\end{defi}

In Example 1 we have two homogeneous Nash configurations, $X^{A}$ and $X^{B}$, 
in which all players play the same strategy, $A$ or $B$ respectively. 

Let us notice that the notion of a Nash configuration is similar to the notion 
of a ground-state configuration in classical lattice-gas models of interacting particles.
We have to identify agents with particles, strategies with types of particles
and instead of maximizing payoffs we should minimize interaction energies.
There are however profound differences. First of all, 
ground-state configurations can be defined only for symmetric matrices; 
an interaction energy is assigned to a pair of particles, payoffs are assigned 
to individual players and may be different for each of them. In fact, it may happen 
that if a player switches a strategy to increase his payoff, the payoff of his opponent 
and of the entire population decreases (like in Prisoner's Dilemma game).
Moreover, ground-state configurations are stable with respect to all local changes,
not just one-site changes like Nash configurations.
It means that for the same symmetric matrix $U$, there may exist a configuration 
which is a Nash configuration but not a ground-state configuration 
for the interaction marix $-U$. The simplest example is given by Example 1
with $a=2, b=c=0$, and $d=1$. $X^{A}$ and $X^{B}$ are Nash configurations but only $X^{A}$ 
is a ground-state configuration for $-U.$  

Games with symmetric payoff matrices are called doubly symmetric 
or potential games \cite{mon}.

More generally, a game is called 
a {\bf potential game} if its payoff matrix can be changed 
to a symmetric one by adding payoffs to its columns. 
As we know, such a payoff transformation does not change strategic 
character of the game, in particular it does not change the set of its
Nash equilibria. More formally, it means that there exists 
a symmetric matrix $V$ called a potential
of the game such that for any three strategies $A, B, C \in S$
\begin{equation}
U(A,C)-U(B,C)=V(A,C)-V(B,C).
\end{equation} 

It is easy to see that every game with two strategies 
has a potential $V$ with $V(A,A)=a-c$, $V(B,B)=d-b$,
and $V(A,B)=V(B,A)=0.$ It follows that an equilibrium is risk-dominant 
if and only if it has a bigger potential.

For players on a lattice, for any $X \in \Omega_{\Lambda}$,
\begin{equation}
V(X)=\sum_{(i,j) \subset \Lambda} V(X_{i},X_{j})
\end{equation}
is then a potential of the configuration $X$.

For any classical lattice-gas model there exists at least one 
ground-state configuration. This can be seen in the following way.
We start with an arbitrary configuration. If it cannot be changed locally
to decrease its energy it is already a ground-state configuration.
Otherwise we may change it locally and decrease the energy of the system.
If our system is finite, then after a finite number of steps we arrive at a
ground-state configuration; at every step we decrease the energy of the system
and for every finite system its possible energies form a finite set.
For an infinite system, we have to proceed ad infinitum converging
to a ground-state configuration (this follows from the compactness of $S^{Z^{2}}$).
Game models are different. It may happen that a game with a nonsymmetric 
payoff matrix may not posess a Nash configuration. The classical example is that 
of the Rock-Scissors-Paper game given by the following matrix.

\vspace{5mm}

\noindent {\bf Example 2}

\hspace{23mm} R  \hspace{2mm} S \hspace{2mm} P  

\hspace{15mm} R  \hspace{3mm} 1  \hspace{3mm} 2 \hspace{3mm} 0

U = \hspace{6mm} S \hspace{3mm} 0  \hspace{3mm} 1 \hspace{3mm} 2

\hspace{15mm} P \hspace{3mm} 2  \hspace{3mm} 0 \hspace{3mm} 1

One may show that this game does not have any Nash configurations on ${\bf Z}$
and ${\bf Z}^{2}$ but many Nash configurations on the triangular lattice.

In short, ground-state configurations minimize the total energy of a particle system, 
Nash configurations do not necessarily maximize the total payoff of a population of agents.

\section{Stochastic Stability}
We describe now the deterministic dynamics of the {\bf best-response rule}. 
Namely, at each discrete moment of time $t=1,2,...$, a randomly chosen player may update 
his strategy. He simply adopts the strategy, $X_{i}^{t}$, which gives him 
the maximal total payoff $\nu_{i}(X_{i}^{t}, X^{t-1}_{-i})$ 
for given $X^{t-1}_{-i}$, a configuration of strategies 
of remaining players at the time $t-1$. 

Now we allow players to make mistakes with a small probability, 
that is to say they may not choose best responses. We will discuss two types 
of such stochastic dynamics. In the first one, the so-called 
{\bf perturbed best response}, a player follows 
the best-response rule with probability $1-\epsilon$ 
(in case of more than one best-response strategy he chooses 
randomly one of them) and with probability $\epsilon$
he makes a ``mistake'' and chooses randomly one of the remaining strategies.
The probability of mistakes (or the noise level) is state-independent here.

In the {\bf log-linear rule}, the probability of chosing by the $i-$th player
the strategy $X_{i}^{t}$ at the time $t$ decreases with the loss of the payoff 
and is given by the following conditional probability:

\begin{equation}
p_{i}^{\epsilon}(X_{i}^{t}|X_{-i}^{t-1})=
\frac{e^{\frac{1}{\epsilon}\nu_{i}( X_{i}^{t},X_{-i}^{t-1})}}{\sum_{Y_{i} \in S}
e^{\frac{1}{\epsilon}\nu_{i}(Y_{i},X_{-i}^{t-1})}},
\end{equation}

Let us observe that if $\epsilon \rightarrow 0$, 
$p_{i}^{\epsilon}$ converges pointwise to the best-response rule.
Both stochastic dynamics are examples of irreducible Markov chains 
(there is a nonzero probability to move from any state to any other state 
in a finite number of steps) with $|S^{\Lambda}|$ states. 
Therefore, they have unique stationary 
probability distributions denoted by $\mu_{\Lambda}^{\epsilon}$. 

The following definition was introduced by Foster and Young \cite{foya}:

\begin{defi}
$X \in \Omega_{\Lambda}$ is {\bf stochastically stable} 
if $\lim_{\epsilon \rightarrow 0}\mu_{\Lambda}^{\epsilon}(X) >0.$
\end{defi} 
If $X$ is stochastically stable, then the frequency of visiting $X$ converges to
a positive number along any time trajectory almost surely. It means that in the long run 
we observe $X$ with a positive frequency.

Stationary distributions of log-linear dynamics can be explicitly constructed 
for potential games. It can be shown \cite{young2} that the stationary distribution
of the log-linear dynamics in a game with the potential $V$ is given by

\begin{equation}
\mu^{\epsilon}_{\Lambda}(X)=\frac{e^{\frac{1}{\epsilon}V(X)}}
{\sum_{Y \in \Omega_{\Lambda}}e^{\frac{1}{\epsilon}V(Y)}}.
\end{equation}

We may now explicitly perform the limit $\epsilon \rightarrow 0$ in (5).
In Example 1, $X^{B}$ has the biggest potential 
(which is equivalent to the risk dominance of $B$) 
so $\lim_{\epsilon \rightarrow 0}\mu_{\Lambda}^{\epsilon}(X^{B})=1 $  
hence $X^{B}$ is stochastically stable (we also say that $B$ is stochastically stable).

Let us now consider coordination games with three strategies and three
symmetric Nash equilibria: $(A,A), (B,B)$, and $(C,C)$. One may say
that $A$ risk dominates the other two strategies if it risk dominates them
in pairwise comparisons. Of course it may happen that $A$ dominates $B$, 
$B$ dominates $C$, and finally $C$ dominates $A$. But even if we do not have
such a cyclic relation of dominance, a strategy which is pairwise risk-dominant 
may not be stochastically stable as we will see below. A more relevant notion seems to be that 
of a global risk dominance \cite{mar}. We say that $A$ is globally risk dominant 
if it provides a maximal payoff against a mixed strategy (a probability distribution
on strategies) which assigns the probability $1/2$ to $A$.
It was shown that a globally risk-dominant strategy is stochastically stable
for some spatial games with nearest-neighbor interactions \cite{ellis1,ellis2}.
A different criterion for stochastic stability was developed by Blume \cite{blume1}. 
He showed (using methods of statistical mechanics) that in games with $k$ strategies 
$A_{i}, i=1,...,k$ and $k$ symmetric Nash equilibria, $A_{1}$ is stochastically stable if
\begin{equation}
\min_{n>1}(U(A_{1},A_{1})-U(A_{n},A_{1})) > \max_{n>1}(U(A_{n},A_{n})-U(A_{1},A_{n})).
\end{equation}
We may observe that if $A_{1}$ satisfies the above condition, then it is pairwise
risk dominant. 

\section{Example}
Let us now present our example of a game with three strategies.
Players are located on a finite subset of the one-dimensional lattice 
${\bf Z}$ and interact with their nearest neighbors only. 
Denote by $n$ the number of players, For simplicity 
we will assume periodic boundary conditions,
that is we will identify the $n+1$-th
player with the first one. In other words, 
the players are located on the circle.

The payoffs are given by the following  matrix:

\noindent {\bf Example 3}
\vspace{2mm}

\hspace{29mm} A \hspace{5mm} B \hspace {4mm} C

\hspace{15mm} A \hspace{5mm} $1+\alpha$  \hspace{2mm} 0 \hspace{3mm} 1.5

U = \hspace{6mm} B \hspace{9mm} 0  \hspace{6mm} 2 \hspace{4mm} 0

\hspace{15mm} C \hspace{9mm} 0  \hspace{6mm} 0 \hspace{4mm} 3
\vspace{5mm}

with $\alpha <0.5$.

\noindent As before, we have three homogeneous Nash configurations, $X^{A}, X^{B}$,
and $X^{C}$. We will consider here both the log-linear and perturbed best-response 
dynamics. The game is not a potential one so there is no explicit formula 
for the stationary distribution. 

To find stochastically stable states, we must resort to different methods. 
We will use a tree representation of the stationary distribution 
of Markov chains \cite{freiwen1,freiwen2,shub}. Let $(\Omega,P)$ be an irreducible Markov chain 
with a state space $\Omega$ and transition probabilities given by 
$P: \Omega \times \Omega \rightarrow [0,1]$. It has a unique stationary distribution. 
A stationary distribution is an eigenvector of the transition matrix $P$ corresponding 
to the eigenvalue 1, i.e., a solution of a system of linear equations. After a specific
rearrangement one can arrive at an expression for the stationary state which involves only
positive terms. This will be very usefull in describing asymptotic behavior of a stationary state. 

For $X \in \Omega$, let X-tree be a directed graph 
on $\Omega$ such that from every $Y \neq X$ there is a unique path to $X$
and there are no outcoming edges out of $X$. Denote by $T(X)$ the set of all X-trees and let 
\begin{equation}
q(X)=\sum_{d \in T(X)} \prod_{(Y,Y') \in d}P(Y,Y'),
\end{equation}
where the product is with respect to all edges of $d$.
 
The following representation of the stationary distribution $\mu$ 
was provided by Freidlin and Wentzell in \cite{freiwen1,freiwen2} (cf also\cite{shub}):
\begin{equation}
\mu(X)=\frac{q(X)}{\sum_{Y \in \Omega}q(Y)}
\end{equation}
for all $X \in \Omega.$

The above characterisation of the stationary distribution
was used to find stochastically states in nonspatial \cite{kmr,young1}
and spatial games \cite{ellis1,ellis2}. Here we will apply it for 
our nonpotential game.

Let us note that $X^{A}$, $X^{B}$, and $X^{C}$ are the only absorbing states 
of the noise-free dynamics. When we start with any state different 
from $X^{A}$, $X^{B}$, and $X^{C}$, then after a finite number of steps 
of the best-response dynamics we arrive at either $X^{A}$, $X^{B}$ or $X^{C}$
and then stay there forever. It follows from the tree representation 
of the stationary distribution that any state different from absorbing states 
has zero probability in the stationary distribution in the zero-noise limit. 
Moreover, in order to study the zero-noise limit of the stationary distribution, 
it is enough to consider paths between absorbing states. More precisely, 
we construct X-trees with absorbing states  as vertices; the family of such 
$X$-trees is denoted by $\tilde{T}(X)$. Let 
\begin{equation}
q_{m}(X)=max_{d \in \tilde{T}(X)} \prod_{(Y,Y') \in d}\tilde{P}(Y,Y'),
\end{equation}
where $\tilde{P}(Y,Y')= max \prod_{(W,W')}P(W,W')$,
where the product is taken along any path joining $Y$ with $Y'$ and the maximum 
is taken with respect to all such paths. 
Now we may observe that if $lim_{\epsilon \rightarrow 0} q_{m}(X^{i})/q_{m}(X^{C})=0,$
$i=A, B$, then $X^{C}$ is stochastically stable. Therefore we have to compare  
trees with the biggest products in (9); such trees we call maximal.

We begin with a stochastic dynamics 
with a state-independent noise. Let us consider the case of $\alpha <0.5$. 
It is easy to see that $q_{m}(X^{C})$ is of order $\epsilon^{2}$,
$q_{m}(X^{B})$ is of order $\epsilon^{n}$, and
$q_{m}(X^{A})$ is of order $\epsilon^{2(n-1)}$. 
We obtained the following theorem.

\begin{theo}
If $\alpha< 0.5$, then $X^{C}$ is stochastically stable in the perturbed best-response dynamics. 
\end{theo} 

Let us now consider the log-linear rule.

\begin{theo}
If $n<2+1/(0.5-\alpha)$, then $X^{B}$ is stochastically stable and
if $n>2+1/(0.5-\alpha)$, then $X^{C}$ is stochastically stable in the log-linear dynamics. 
\end{theo} 

{\bf Proof}: The following are maximal A-tree, B-tree, and C-tree:
$$B \rightarrow C \rightarrow A, \hspace{3mm} C \rightarrow A \rightarrow B,
\hspace{3mm} A \rightarrow B \rightarrow C,$$

where the probability of $A \rightarrow B$ is equal to
\begin{equation}
\frac{1}{1+1+e^{\beta(2+2\alpha)}}(\frac{1}{1+e^{-2\beta}+e^{\beta(-1+\alpha)}})^{n-2}
\frac{1}{1+e^{-4\beta}+e^{-4\beta}},
\end{equation}
the probability of $B \rightarrow C$ is equal to
\begin{equation}
\frac{1}{1+1+e^{4\beta}}(\frac{1}{1+e^{-\beta}+e^{-1.5\beta}})^{n-2}\frac{1}
{1+e^{-6\beta}+e^{-3\beta}},
\end{equation}
and the probability of $C \rightarrow A$ is equal to
\begin{equation}
\frac{1}{1+e^{-3\beta}+e^{3\beta}}(\frac{1}{1+e^{-\beta(2.5+\alpha)}
+e^{\beta(0.5-\alpha)}})^{n-2}\frac{1}{1+e^{-2\beta(1+\alpha)}+e^{-2\beta(1+\alpha)}}
\end{equation}

Let us observe that 
\begin{equation}
P_{B \rightarrow C \rightarrow A}= O(e^{-\beta(7+(0.5-\alpha)(n-2))}),
\end{equation}
\begin{equation}
P_{C \rightarrow A \rightarrow B}=
O(e^{-\beta(5+2\alpha+(0.5-\alpha)(n-2))}),
\end{equation}
\begin{equation}
P_{A \rightarrow B \rightarrow C}= O(e^{-\beta(6+2\alpha)}),
\end{equation}
where $\beta = 1/\epsilon$ and $\lim_{x \rightarrow 0}O(x)/x =1.$

Now if $n<2+1/(0.5-\alpha)$, then 
\begin{equation}
\lim_{\epsilon \rightarrow 0}\frac{q_{m}(X^{C})}{q_{m}(X^{B})}=
\lim_{\epsilon \rightarrow 0}\frac{P_{A \rightarrow B \rightarrow C}}
{P_{C \rightarrow A \rightarrow B}}=0
\end{equation}
which finishes the proof.
\vspace{2mm}

It follows that for small enough $n$, $X^{B}$ is stochastically stable
and for big enough $n$, $X^{C}$ is stochastically stable. 
We see that adding more players to the population 
may change the stochastic stability of Nash configurations. 
Let us also notice that the strategy $C$ is globally risk dominant. 
Nevertheless, it is not stochastically stable in the log-linear dynamics 
for a sufficiently small number of players. 

Let us now discuss the case of $\alpha=0.5.$

\begin{theo}
If $\alpha=0.5$, then $X^{B}$ is stochastically stable for any $n$ in the log-linear dynamics. 
\end{theo}

{\bf Proof}: 
$$\lim_{\epsilon \rightarrow 0}\frac{q_{m}(X^{C})}{q_{m}(X^{B})}=
\lim_{\epsilon \rightarrow 0}\frac{e^{-4\beta}e^{-3\beta}}{(1/2)^{n-2}e^{-3\beta}e^{-3\beta}}=0.$$
\vspace{2mm}

$X^{B}$ is stochastically stable which means that for any fixed number of players, 
if the noise is sufficiently small, then in the long run we observe $B$ players 
with an arbitrarily high frequency. However, for any low but fixed noise, 
if the number of players is big enough, the probability of any individual configuration 
is practically zero. It may happen though that the stationary distribution 
is highly concentrated on an ensemble consisting of one Nash configuration 
and its small perturbations, i.e. configurations, where most players play the same strategy.
We will call such configurations {\bf ensemble stable} \cite{statmech}.
An ensemble-stable configuration may not be stochastically stable. 
In fact, we expect that $X^{C}$ is ensemble stable because its lowest-cost excitations 
occur with a probability of order $e^{-3\beta}$ and those from $X^{B}$  
with a probability of order $e^{-4\beta}$.  We observe this in simple Monte-Carlo 
simulations. 

\section{Discussion}
We studied the problem of equilibrium selection in spatial games
with many players. We showed that stochastic stability of Nash configurations
may depend on the number of players. In particular, 
we presented an example with the globally risk dominant
equilibrium which is stochastically stable in the perturbed best response
dynamics and is not stochastically stable in the log-linear one 
if the number of players is smaller than some critical value
which grows to infinity when some payoff parameter approaches a critical value. 
\vspace{2mm}

\noindent {\bf Acknowledgments}: I thank Christian Maes and Joseph Slawny
for useful conversations and Rudolf Krejcar for writing Monte-Carlo programs
to simulate stochastic dynamics of spatial games. 
Financial support by the Polish Committee for Scientific Research
under the grant KBN 5 P03A 025 20 is kindly acknowledged.

\end{document}